\documentclass[twocolumn,showpacs]{revtex4}

\usepackage{graphicx}

\begin{document}

\newcommand{\xfel}{\textsc{xfel}}

\title{Ionization of clusters in strong X-ray laser pulses}
\author{Ulf Saalmann}
\author{Jan-Michael Rost}
\affiliation{Max-Planck-Institut f{\"u}r Physik komplexer Systeme,
  N{\"o}thnitzer Str.\ 38, 01187 Dresden, Germany}
\date{\today}

\begin{abstract}\noindent
  The effect of intense X-ray laser interaction on argon
  clusters is studied theoretically with a mixed
  quantum/classical approach.
  In comparison to a single atom we find that ionization of
  the cluster is suppressed, 
  which is in striking contrast to the observed behavior of
  rare-gas clusters in intense optical laser pulses.
  We have identified two effects responsible for this
  phenomenon: 
  A high space charge of the cluster in combination with a small
  quiver amplitude and delocalization of electrons in the
  cluster.
  We elucidate their impact for different field strengths and
  cluster sizes. 
\end{abstract}


\maketitle

The advent of femtosecond laser pulses has triggered 
numerous activities in the field of laser-matter interaction.
In particular novel, non-linear processes in atoms
induced by the available \emph{high intense\/} fields 
from these ultrashort pulses have been a challenge 
for experiment and theory \cite{pi01}.
To understand more generally the energy transfer 
from intense laser light to matter, 
complex targets like atomic clusters have been studied
\cite{diti+97,shdi+96,mcth+94,snbu+96}.
These experiments received considerable attention due to 
the observed dramatic effects 
like emission of very fast ions \cite{diti+97}
and electrons \cite{shdi+96} or the
production of coherent X-ray radiation \cite{mcth+94}.
So far such studies are almost exclusively restricted to visible
or infrared wavelengths \cite{xenon}. 
The new X-ray free electron laser (\xfel) sources,
under construction at DESY in Hamburg \cite{xfel} 
and at the LCLS in Stanford \cite{ar02}, will change this situation.
They can deliver intense laser pulses at \emph{high frequencies\/}
(from VUV to hard X-ray)
and thus open a new regime of strong-field atomic physics.

Here we present theoretical investigations of X-ray 
(frequency $\hbar\omega=350\,\mbox{eV}$, i.\,e.\
wavelength $\lambda\approx3.5\,\mbox{nm}$)
laser interaction with argon clusters at intensities
of $I\approx3.5{\cdot}10^{14}\ldots3.5{\cdot}10^{18}$\,W/cm$^2$.
In this laser regime 
the interaction is notably different 
from long-wavelength pulses which is evident 
 from the ponderomotive energy 
$E_\mathrm{pond}\sim I/\omega^2$. It represents
the average kinetic energy of a free electron in a laser field,
while $\Delta x \sim \sqrt I/\omega^2$ is the quiver amplitude
of the electron, 
i.\,e., the spatial excursion in the laser field.
For the given laser parameters one finds
$E_\mathrm{pond}\approx 0.4\,\mbox{meV}\ldots 4\,\mbox{eV}$,
vastly different from 
$E_\mathrm{pond}\approx 20\,\mbox{eV}\ldots 200\,\mbox{keV}$
for a 780-nm-laser at the same intensities.
The  small ponderomotive energy has two 
important consequences:
(i)~Despite the high intensities the laser-atom interaction 
is of non-relativistic and perturbative nature.
The latter is also clear from the so-called Keldysh parameter 
$\gamma$ which gives the ratio between the tunneling time
and the laser period \cite{ke65}
and can be rewritten in terms of the binding energy 
$E_\mathrm{bind}$ and the ponderomotive energy $E_\mathrm{pond}$
as $\gamma=(E_\mathrm{bind}/2E_\mathrm{pond})^{1/2}$.
For the $L$ and $M$-shells of argon
with $E_\mathrm{bind}=326,\:249,\:29.3,\:15.8$\,eV 
we always find  $\gamma>1$ 
and for the inner shells even $\gamma\gg1$.
Obviously the laser period is far too small for 
field- or tunnel-ionization.
Rather, ionization is due to single-photon absorption. 
(ii)~Ponderomotive effects \cite{bufr+87} are completely negligible
with quiver amplitudes of $\Delta x \approx0.0003\ldots0.03$\,\AA.  
 
Under the aforementioned laser conditions one expects exhaustive
ionization leading subsequently to a gigantic Coulomb explosion
of the irradiated argon clusters, 
since ionization proceeds fundamentally different from that
under optical laser impact. 
Ionization starts from the inside because
photoionization cross sections at X-ray wavelengths 
are considerably higher for the inner shells  than for the valence
shells \cite{am90}.
Typically the inverse rates are with
$1\ldots10\,\mbox{fs}$ \cite{rate} 
much smaller than the pulse length of about
$100\,\mbox{fs}$. 
Hence, multiple single-photon ionization is possible, in
particular because the inner-shell holes created by
photoionization are refilled by Auger-like processes. 
The Auger decay is more or less independent of the atomic charge
state and occurs fast,
typical times are $0.2\ldots5\,\mbox{fs}$ \cite{kosu+95}.
Due to this almost instantaneous refilling of 
the inner shells they can be ionized many times during the
pulse and thus the atoms can be efficiently ``pumped dry''.
This occurs ``inside-out'' and is the exact opposite to the
ionization mechanism in the visible wavelengths regime where the
electrons are removed from the outside like shells of an onion
\cite{rosc+97}. 
Finally, local ionic charges in the cluster, generated in the course
of the ionization process, may be screened by the weakly bound
valence electrons of cluster atoms.
Therefore, these electrons are allowed to tunnel through the
barriers between the mother atom and neighboring ions.
Such electrons will predominantly move \emph{inside\/} the
cluster since there is no field 
(such as the quasi-static electric field of low-frequency lasers) 
which would drive them away from the charged cluster into the
continuum. 

In order to gain insight into the laser induced dynamics we use a
mixed quantum/classical approach  similar to those
successfully applied in studying rare-gas clusters in strong pulses of
visible light \cite{rosc+97,isbl00,siro01}.
Electrons initially bound to atoms in the cluster are not explicitly 
treated, they only enter through their binding energy.
The ionization from the mother atom in the cluster, also called inner 
ionization, is modelled statistically by sudden transitions according to
quantum mechanical rates. An inner ionization event gives birth
to a quasi-free electron which is subsequently 
propagated classically along with the ions and other quasi-free electrons
with all Coulomb forces included. 
Note that a quasi-free electron is not necessarily  ionized 
with respect to the cluster as a whole, quasi-free means only inner-ionized 
with respect to the mother atom or ion.

The ionization/excitation dynamics
in strong X-ray pulses is considerably more complicated
than field \cite{rosc+97} or tunnel \cite{isbl00,siro01} 
ionization in optical pulses
because inner-shell electrons and intra-atomic processes
are involved.
For the photoionization rates we use a parameterization of the cross
section which covers all charge states of atomic argon \cite{veya+93}. 
For high ionic charges the electronic binding energy can become so
large that single photoionization is impossible.  Non-dipole effects
do not have any crucial impact apart from distortions of the angular
distribution of the photoelectrons \cite{co93}.  For modelling the
intra-atomic decay cascades we use direct Hartree-Fock calculations
\cite{kosu+95}.  They provide branching ratios and decay rates for
Auger, Coster-Kroning, and radiative transitions.  The latter ones
are typically at least one order of magnitude slower \cite{kosu+95}. 
For the intra-cluster charge equilibration we calculate tunnel (or
over-the-barrier) rates in analogy to \cite{isbl00,siro01}.  All the
rates are used to decide at every time step of the simulation whether
a particular transition occurs or not.  Once the electrons are created
with their respective kinetic energy (to guarantee energy conservation
in the photo-absorption process or in the Auger decay \cite{levels})
they are propagated in the field of the other particles.

\begin{figure}[t]
  \centering
  \includegraphics[width=70mm]{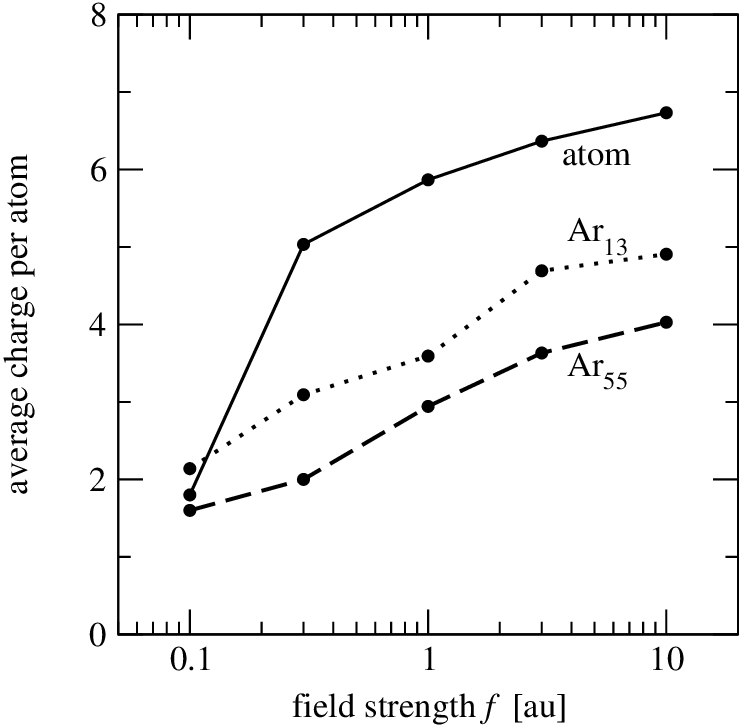}
  \caption{Average charge per atom for two cluster sizes
    Ar$_{13}$ and Ar$_{55}$ produced by an \xfel\ pulse 
    ($\hbar\omega=350$\,eV, $T=100$\,fs)
    as a function of the field strength $f$
    compared to atomic argon as target.
    The points are averages from 10 simulations.}
  \label{fig:fich}
\end{figure}

From the physics described so far it is clear that the initial 
coupling to the laser is individual atom-photon interaction only.
Hence, it is particularly interesting to see how the
cluster environment affects the laser interaction which can be
most clearly seen from calculations comparing atoms and clusters. 
The laser pulse is given by
$f(t) = f\cdot s(t)\cdot\sin(\omega t)$
with the electric field strength~$f$,  the laser frequency~$\omega$, 
and the  pulse shape function
$$
s(t) =\left\{\begin{array}{lll}
    \sin^2\left(\frac{\pi}{2}\:\frac{t}{T/5}\right) 
    & 0\le t\le T/5,\\
    1 & T/5\le t\le 4T/5, \\
    \sin^2\left(\frac{\pi}{2}\:\frac{T-t}{T/5}\right) \quad
    & 4T/5\le t\le T,\\
    0 & \mbox{otherwise,}
  \end{array}\right.
$$
where $T$ is the pulse length.
Figure \ref{fig:fich} shows the final charge per atom for two
different clusters, Ar$_{13}$ and Ar$_{55}$, compared to a single
atom for a laser pulse
with $\hbar\omega=350$\,eV, $T=100$\,fs, and $f=0.1\ldots10$\,au
(atomic unit).
At  lower field strengths ($f=0.1$\,au), only a single
photoionization event with one subsequent Auger decay per atom
occurs independently of the cluster size.
After a quite steep rise  the final charge  starts to saturate
in the atomic case for stronger fields ($f>0.3$\,au) 
due to the  fact that  single photoionization becomes impossible 
beyond a certain charge state of the ion at the given photon 
frequency.
The rise of the final charge per atom in the cluster is notably
weaker. 
Therefore 
--- in striking contrast to optical laser-cluster interaction
--- clusters are less effectively ionized at  high fields
than atoms.
The reduction is even more pronounced for the larger cluster
(cf.\ Fig.~\ref{fig:fich}).
One reason for the reduced ionization in the cluster compared to the 
atom is the much larger space charge produced in a cluster.
At $f=10$\,au the total cluster charges $Q$ after the pulse
are $Q_1\approx7$, $Q_{13}\approx65$, and $Q_{55}\approx220$
for atoms and clusters of 13 and 55 atoms, respectively.
\begin{figure}[b]
  \centering
  \includegraphics[width=70mm]{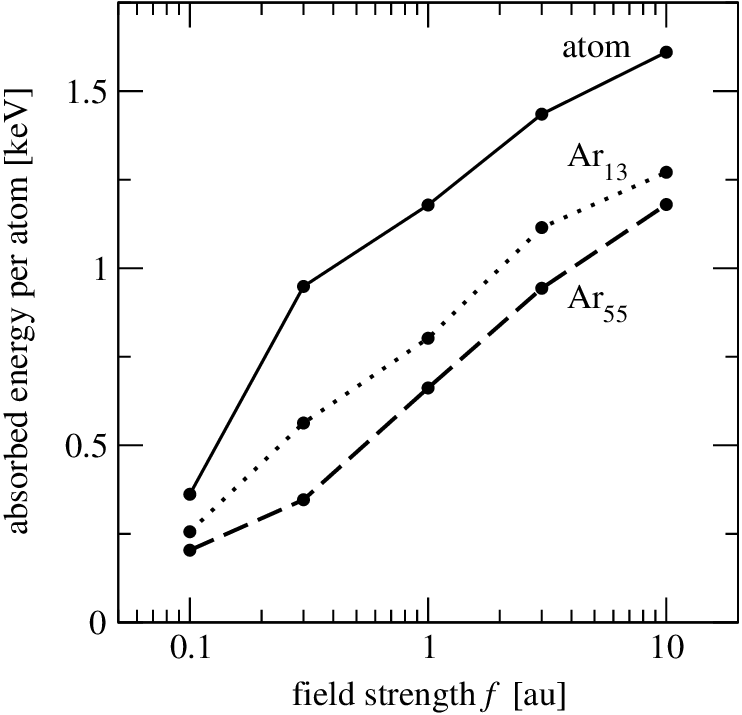}
  \caption{Absorbed energy per atom 
    as a function of the field strength $f$
    for the three targets and the laser pulse shown in
    Fig.~\ref{fig:fich}.}
  \label{fig:aben}
\end{figure}%
Such high space charges suppress ionization, 
because the absorption of one single photon transfers only a
fixed amount of energy to the electron.
In our model, these post-collisional interaction effects are
taken into account by the propagation of the photoionized electrons. 
For larger cluster and higher fields,
i.\,e.\ higher space charges, one finds that
an increasing number of electrons are bound at the end of the
pulse to one of the fragment ions.

Figure~\ref{fig:aben} shows  the energy
absorbed from the laser pulse for the same clusters.
This observable provides direct insight into the photoionization
since all the other processes
(intra-atomic decay, intra-cluster screening)
are not influenced by the laser due to its high frequency.
Surprisingly, for all field strengths considered, the absorption of
photons itself is reduced in the environment provided by the cluster.
This was unexpected because of the fact that predominantly
deep-lying inner-shell electrons are affected
and possible effects on the photoionization
rates of these strongly localized electrons
from neighboring ions are unlikely and in fact 
not contained in the model. 

\begin{figure}[b]
  \centering
  \includegraphics[width=80mm]{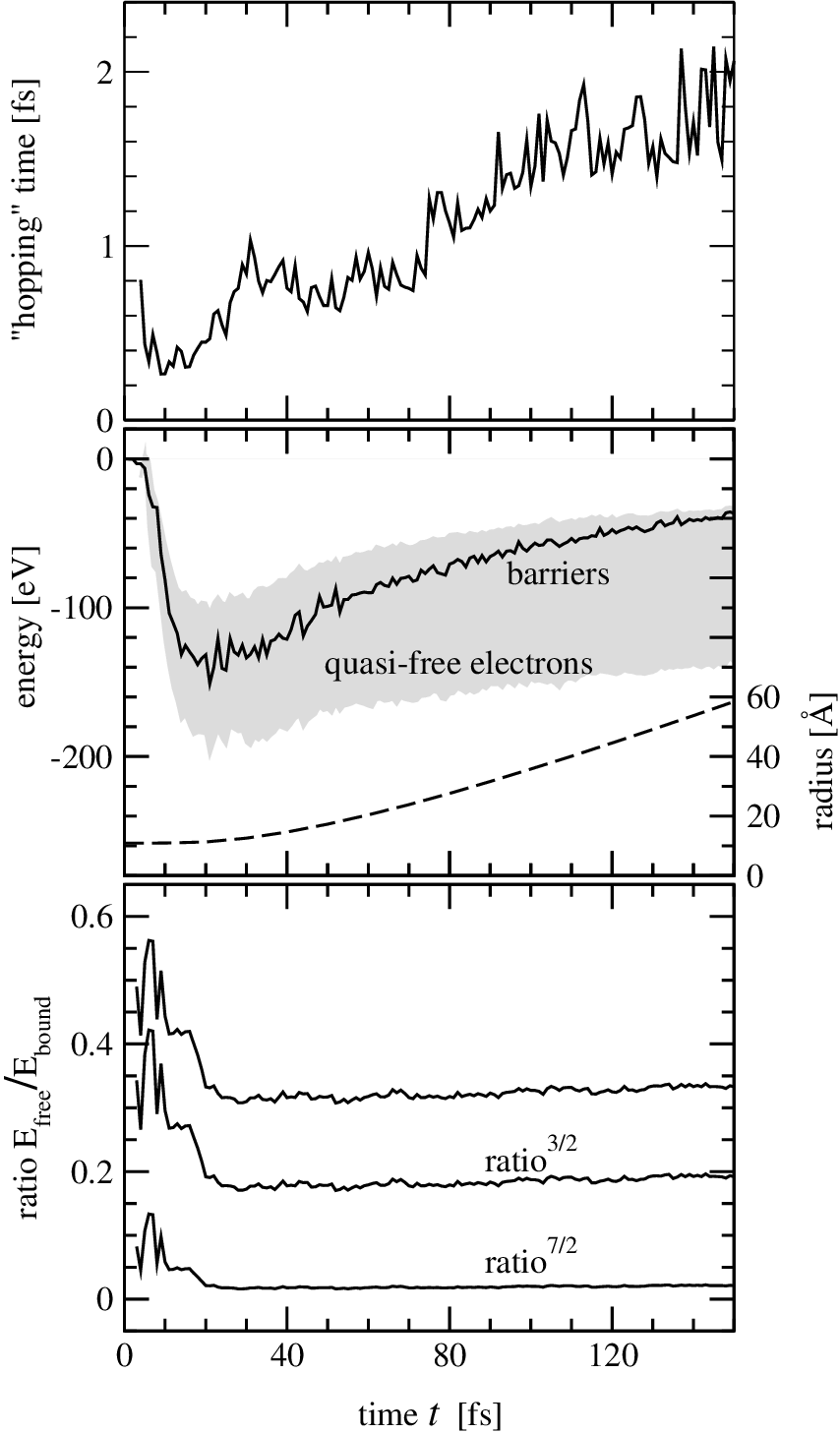}
  \caption{Time evolution of an Ar$_{55}$ cluster
          in an \xfel\ pulse with $\hbar\omega=350$\,eV, 
          $f=1$\,au, and $T=100$\,fs.
          {\em Upper panel:\/} Average time a quasi-free electron
          stays at a particular cluster atom.
          {\em Middle panel:\/} Energy range of quasi-free electrons
          (mean value $\pm$ standard deviation) compared to
          the average of the barriers between ions in the
          cluster.
          {\em Dashed line:\/}
          Radius of the cluster, see right axis.
          {\em Lower panel:\/} Average of ratios 
          $E_\mathrm{free}/E_\mathrm{bound}$ and powers of it, see text.}
        \label{fig:te55}
\end{figure}

The time evolution of the cluster dynamics reveals that a
delocalization of the valence electrons is indirectly responsible. 
It has mainly two effects: 
Firstly, the photoionization cross sections become very small
since the electrons are far away from the nucleus.
Secondly, also the Auger decay rates are reduced because the
overlap with the core holes becomes smaller. 
Before quantifying these effects we will look at the
electron dynamics in detail.
Figure~\ref{fig:te55} shows the 
time evolution of an Ar$_{55}$ cluster for the same pulse
parameters as before at a field strength of $f=1$\,au.
At every time step each electron is assigned to the 
cluster ion with which it has the largest instant  binding energy.
In this way we define a ``hopping'' time which is the time an
electron stays at the same ion before it moves  to another
one. 
The upper panel of Fig.~\ref{fig:te55} shows that the average
``hopping'' time is less than 1\,fs  
during the first half of the pulse ($t<50$\,fs).
This is very short compared to typical inverse ionization rates
and indicates that the electrons move almost freely
inside the cluster volume. Hence, the low barriers
 between
the cluster ions in the early phase of the pulse make inner 
ionization very effective.
As can be seen in the middle panel of Fig.~\ref{fig:te55},
a considerable fraction of the quasi-free electrons 
has energies above these barriers.
This will change only towards the end of the pulse ($t=100$\,fs)
 when due to the expansion of the cluster
(cf.\ the cluster radius in  Fig.~\ref{fig:te55}) 
the barriers rise again.

It is difficult to assign photoionization or autoionization
rates to this ``sea'' of electrons.
However, one can give estimates of these rates using the fact that 
they depend in a characteristic manner on the energy of the 
respective bound electron. The quasi-free electrons have a binding energy $E_\mathrm{free}$
(since they are still bound with respect to the full cluster). 
This energy will be compared to $E_\mathrm{bound}$ the binding
energy of the weakest bound electron in the atom, 
which is among the bound electrons the most stable one against
ionization. 
The ratios $E_\mathrm{free}/E_\mathrm{bound}$, averaged over all the
electrons, are shown in the lower panel of Fig.~\ref{fig:te55}. 
Compared to the low rates of the bound electrons
the photoionization rate falls off like 
$(E_\mathrm{free}/E_\mathrm{bound})^{7/2}$ \cite{am90} 
and the Auger decay rate roughly like  
$(E_\mathrm{free}/E_\mathrm{bound})^{3/2}$ \cite{po88}.
Due to this scaling the rates are reduced  during
the pulse by a factors of about 5 and 20, respectively.
Obviously, also those electrons with energies below
the barriers are still well above the highest bound electrons
and hence fairly delocalized.
One has to emphasize that this estimate is an upper bound since the
rates decrease even more  for higher angular
momentum states \cite{am90,po88}
which are likely to be populated by the intra-cluster dynamics. 
Therefore, we regard it as save to neglect absorption of photons or
autoionization of the quasi-free electrons.

As we have seen in addition to simply lower
absorption rates the reduction of  photon absorption
is also due to the fact that the inner shells to be ionized are 
no longer efficiently  refilled by inter-atomic decay, i.\,e. the
cluster is temporarily hollow. 
In order to assess the relative importance of this effect
compared to the suppression of  ionization due to the cluster
space charge we have repeated our cluster calculations without the 
possibility for tunneling of electrons between cluster ions.
\begin{figure}[t]
  \centering
  \includegraphics[width=80mm]{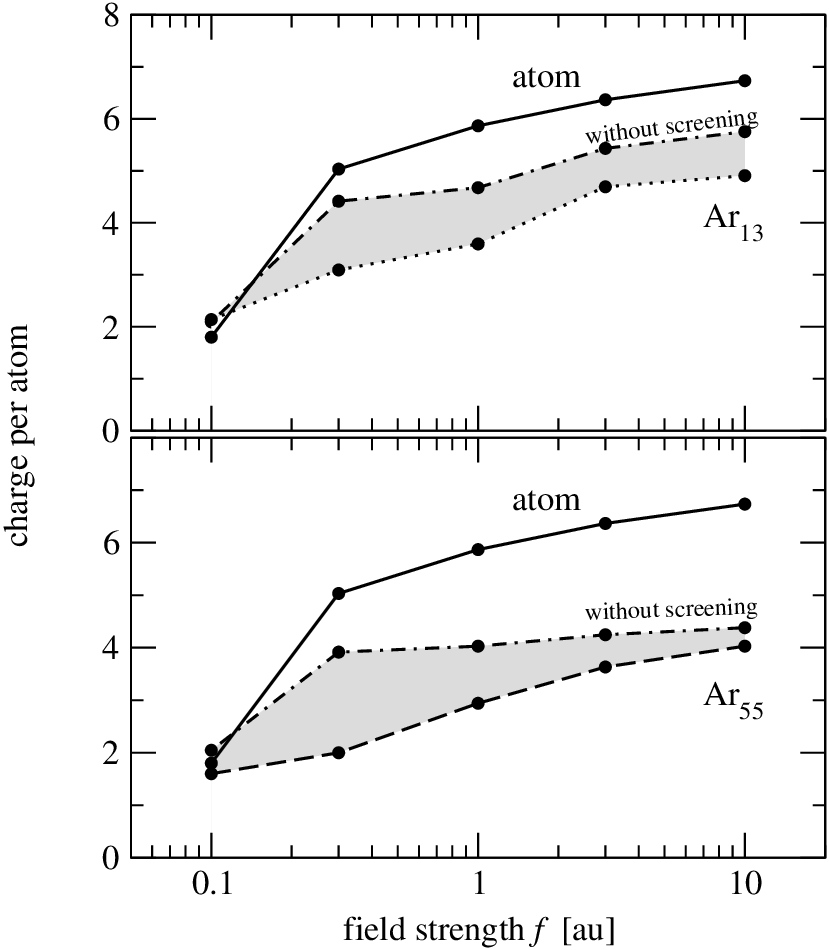}
  \caption{Average charge per atom for two cluster sizes
    Ar$_{13}$ and Ar$_{55}$ produced by an \xfel\ pulse
    (same parameters as in Fig.~\ref{fig:fich}) 
    as a function of the field strength $f$.
    {\em Dot-dashed line:} Restricted cluster calculation where
    intra-cluster screening was precluded.}
  \label{fig:fich2}
\end{figure}
Figure~\ref{fig:fich2} shows the final charges from these
restricted calculations for Ar$_{13}$ and  Ar$_{55}$ 
along with the full calculation for these clusters and the atom.
The difference between the two cluster calculations with and without 
tunneling 
(marked by grey shading in Fig.~\ref{fig:fich2})
accounts  for the delocalization effect.
The difference between the restricted cluster
calculation
(the dash-dotted line in Fig.~\ref{fig:fich2})
and that for the atom reveals the space
charge effect.
For field strengths $f\ge0.3$\,au, where differences between
atom and clusters appear, the space charge effect is initially
weaker. 
This changes for stronger fields:  
Whereas for the smaller cluster Ar$_{13}$ both are of the same
magnitude at $f=10$\,au,
for the larger cluster Ar$_{55}$ the space charge effect
dominates at this field strength.

In conclusion, we have found suppression of ionization for small
argon cluster compared to isolated atoms in the same pulse of
intense X-ray radiation. 
This behavior is in striking contrast to that of rare-gas
clusters in intense optical lasers.
Two effects are responsible for the difference:
Firstly, 
the high positive space charge of the cluster hinders electron emission since 
the space charge is not compensated by a large quiver motion. 
Secondly,
delocalization of electrons in the cluster reduces
photoionization as well as autoionization  drastically.
Both effects are more important for the larger cluster
investigated and the relative weight of both effects depends on field strength
and cluster size. Our findings indicate that in general the coupling of
energy from the laser light to matter is less effective at high frequencies.
This has important consequences for \xfel\ imaging 
applications since it implies a higher
damage threshold. That is, fragile samples might survive irradiation by 
intense pulses of high frequency better than anticipated.

We would like to thank Christian Siedschlag for helpful discussions.


\begin{thebibliography}{10}
\bibitem{rate}
  For a laser with $\hbar\omega=350\,\mbox{eV}$ at
  $I=3.5{\cdot}10^{16}\,\mbox{W/cm}^2$ the photoionization rates for the
  $L$-shells of neutral argon are about $\Gamma_{L_1}=0.03\,\mbox{fs}^{-1}$
  and $\Gamma_{L_{23}}=0.22\,\mbox{fs}^{-1}$ \cite{veya+93}.
\bibitem{levels}
  This requires the calculation of atomic energies for all
  possible electronic configurations. 
  In the case of argon these are 1323.
\end{thebibliography}
\end{document}